\documentclass[12pt]{iopart}

\usepackage{graphicx}
\usepackage{bpchem}
\usepackage{siunitx}
\usepackage{color} 
\usepackage{verbatim}
\usepackage{amssymb}

\def\hDia{6.5cm}

\def\TReg{\textsuperscript{\textregistered}}

\definecolor{cyan1}{rgb}{0, 1, 1}
\definecolor{red1}{rgb}{1, 0, 0}
\definecolor{blue1}{rgb}{0,0,1}
\definecolor{green1}{rgb}{0,1,0}

\begin{document}
\title{Numerical simulation of the Tayler instability in liquid metals}
\author{Norbert Weber, Vladimir Galindo, Frank Stefani, Tom Weier and
  Thomas Wondrak}
\address{Helmholtz-Zentrum Dresden-Rossendorf, 
P.O. Box 510119, 01314 Dresden, Germany}
\ead{Norbert.Weber@hzdr.de}
\begin{abstract}
The electrical current through an incompressible, viscous and 
resistive liquid conductor produces
an azimuthal magnetic field that becomes unstable when the
corresponding Hartmann number exceeds a critical value in the
order of 20. This Tayler instability, which is not only discussed as
a key ingredient of a non-linear stellar dynamo model (Tayler-Spruit
dynamo), but also as a limiting factor for the maximum size 
of large liquid metal batteries, was recently
observed experimentally in a column of a liquid metal 
\cite{Seilmayer2012}.
On the basis of an integro-differential equation approach,
we have developed a fully three-dimensional numerical code,
and have utilized it
for the simulation of the Tayler instability at typical 
viscosities and resistivities of liquid metals. The 
resulting growth rates are in good 
agreement with the experimental data. We illustrate the
capabilities of the code for the detailed simulation 
of liquid metal battery problems in realistic geometries.

\end{abstract}

\submitto{\NJP}
\maketitle

\section{Introduction} 

Current driven instabilities have been known in plasma physics for many
decades \cite{Tayler1960}. A paradigm for their occurrence is the 
$z$-pinch, a cylindrical plasma column with an electrical current 
in direction of the cylinder axis that produces an azimuthal
magnetic field. Recently,  Bergerson \etal  
have evidenced the onset of kink-type instabilities
in a line-tied screw pinch when the
ratio of axial to azimuthal magnetic field 
(the safety parameter) drops below a certain critical value
\cite{Bergerson2006}.

For a purely azimuthal magnetic field, the onset of 
current driven instabilities
depends basically on the detailed radial dependence $B_{\varphi}(r)$.
For the kink-type ($m=1$) instability, the relevant criterion is
\begin{eqnarray}
\frac{\partial (r B^2_{\varphi}(r))}  {\partial r} > 0,
\end{eqnarray}
which had been  derived by Vandakurov \cite{Vandakurov1972} and Tayler 
\cite{Tayler1973}.
Strictly speaking, this condition holds only 
without taking into account the stabilizing role of  rotation, 
viscosity, resistivity, or 
density stratification. Focusing on the latter, 
the term {\it Tayler instability} had been coined 
by Spruit \cite{Spruit2002} 
to describe a situation in which the azimuthal magnetic field 
becomes strong enough to act against the stable stratification 
in a star. This instability is quite remarkable since it 
could provide the second ingredient, in addition to
differential rotation, for an alternative type of stellar dynamos,
which is called now the {\it Tayler-Spruit dynamo}.
Although this sort of non-linear dynamo is presently
controversially discussed and might fail to work 
under realistic conditions \cite{Zahn2007}, 
the underlying kink instability could have
important astrophysical consequences, in particular for the
extreme spin down of the cores of white dwarfs \cite{Suijs2008}, 
for chemical mixing in stars \cite{Yoon2012}, or for the 
occurrence of helical structures in cosmic jets \cite{Moll2008}.  

Resistivity and viscosity have a similar stabilizing effect 
on the kink instability as density stratification.
By slightly stretching the original semantics \cite{Spruit2002}, we 
will use here the terminus {\it Tayler instability} (or TI) for this 
viscous and resistive case, too, keeping in mind that
in either case the azimuthal magnetic field 
has to exceed a certain critical value in order to 
become unstable.

In plasma physics the effect of viscosity and resistivity 
has been studied for various
boundary conditions and profiles of the electrical current 
\cite{Spies1988,Cochran1993}.
Apart from details, it was shown 
that the onset of TI occurs only when
the product of the Lundquist number $S=\mu_0 \sigma R v_A$ 
and the so-called Alfv\'en-Reynolds number $A=\nu^{-1} R v_A$
exceeds a value in the order of 10$^3$. Here,
$\mu_0$ is the magnetic permeability constant, $\sigma$ and $\nu$ are the
conductivity and the viscosity of the fluid, $R$ is the radius 
of the pinch, and
$v_A:=B(\mu_0 \rho)^{-1/2}$ is the Alfv\'en speed which is 
proportional to the
magnetic field $B$ ($\rho$ is the density of the fluid).

Going over from plasma physics, with all its peculiarities due to
compressibility, anisotropic viscosities and conductivities,
as well as complicated boundary conditions,
to the paradigmatic case of an incompressible, viscous 
and resistive cylindrical fluid with a homogeneous current 
distribution,
R\"udiger \etal were able to show that the key
parameter is the Hartmann number, 
$\mathit{Ha}=B R (\sigma/\rho \nu)^{1/2}$, 
which should exceed a value of approximately 25 for the
TI to set in when we take $B=B_\varphi(R)$ \cite{RUEDIGER2011}. This is completely consistent 
with the former results in plasma physics because $\mathit{Ha}^2=S A$.
With typical viscosities and conductivities of liquid metals the
critical electrical current is in the order of $10^3$ \si{\ampere}. 
Using the eutectic alloy GaInSn,
we had experimentally confirmed the onset of TI 
at a critical current
of approximately \SI{2.7}{\kilo\ampere}, as well as the numerically 
predicted 
increase of this
threshold when insulating inner cylinders were inserted into the
liquid metal.

\begin{figure}[hbt]
\centerline{
\includegraphics[height=6cm]{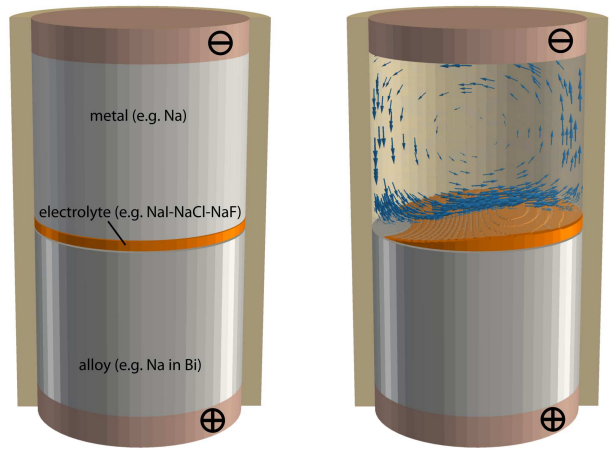}}
\caption{Sketch of a liquid metal battery with typical inventory (left). 
The electrolyte works as ion conductor and separates the two liquid metals. 
A movement of the fluid may wipe the electrolyte and lead to an
internal short-circuit and thereby a battery failure (right). The TI will induce one or several convection 
cells in the conducting liquid metal layers.}
\label{fig:battery}
\end{figure}

Interestingly, currents in this order would indeed be relevant for  
large-scale liquid metal batteries which are presently strongly discussed as cheap means for the storage of highly
intermittent renewable energies \cite{Huggins2010,
  Bradwell2012}. Smaller versions of these devices were already
studied in the sixties \cite{Weaver1962, Cairns1969}.
Basically, liquid metal batteries 
consist of a self-assembling stratified system of a heavy liquid
metalloid (e.g. Bi, Sb) at the bottom, an appropriate
molten salt mixture as electrolyte in the middle, and a light
alkaline or earth alkaline metal (e.g. Na, Mg) at the
top (see figure \ref{fig:battery}). 
Choosing \BPChem{Na} and \BPChem{Bi} as an example, 
\BPChem{Na} will loose 
one electron during the discharge process, turning into 
\BPChem{Na\^+}. This ion diffuses
through the electrolyte into the lower \BPChem{Bi} layer
where it is reduced and alloys with 
\BPChem{Bi} to \BPChem{NaBi}. Thus, the lower part
of the battery will
increase in size during discharge and the volume of the upper part
will decrease correspondingly.

While small versions of such batteries have already
been tested \cite{Bradwell2012}, the occurrence of the TI
could represent a serious problem for 
the integrity of the
stratification in larger batteries \cite{Stefani2011}. 
This is all the more an issue 
as the highly resistive electrolyte should be
chosen as thin as possible in order to maintain a reasonable cell voltage.
In \cite{Stefani2011}  we had proposed a
simple trick to suppress the TI in liquid metal batteries by
just returning the battery current through a bore in the
middle. By the resulting change of the radial dependence
of $B_{\varphi}(r)$ it is possible to prevent the (ideal)
condition (1) for the onset of the TI.

In spite of such attempts to suppress the TI, it is 
worthwhile to study in detail
the flow structure that would arise from it, and the resulting
consequences on the stratification of the three layer system.
A first step in this direction is the 
determination of the final velocity field in the saturated 
state of the TI. The code utilized up to now for the simulation of the
liquid metal experiment
was well capable of determining the critical current and the growth
rates for the onset of the TI \cite{Ruediger2012b}, but the resulting 
scale of the velocity in the saturated state was 
less secure. This has to do with the 
fact that liquid metals
are characterized by a very small magnetic Prandtl number 
$Pm:=\nu \mu_0 \sigma\sim 10^{-6}...10^{-5}$. 
The usual numerical schemes for the simulation 
of TI, which solve the
Navier-Stokes equation for the velocity and the
induction equation for the magnetic field, are working 
typically only for values of  $Pm$ down to 10$^{-2}$.
When applying the code at these much too high $Pm$,
and scaling the resulting velocity level by $\nu/R$, 
one arrives at velocity scales in the order
of mm/s. However, an extrapolation over 
4 orders of magnitude is somehow risky, and needs definitely a 
justification by codes that can cope with realistic values 
of $Pm$.

In this paper, we circumvent the $Pm$ limitations of 
those codes that rely on the differential equation approach. We do 
this by replacing the solution of the
induction equation for the 
magnetic field by applying the so-called quasistatic 
approximation \cite{Davidson2001}. This approximation means that 
we skip the explicit time dependence of the magnetic field by 
computing the electrostatic potential by a Poisson equation,
and deriving then the electric current density.
However, in contrast to many other applications in which this procedure 
is sufficient,
in our case we cannot avoid to compute the induced magnetic field, too.
The reason for that is the existence of an externally applied current
in the fluid. Computing the Lorentz force term it turns out that
the product of the applied current times the induced field is of the same
order as the product of the magnetic field (due to the applied current) times
the induced current.
Here, we compute the induced current density from the induced magnetic field
by means of Biot-Savart's law. This way we arrive at an 
integro-differential
equation approach, as it had already been used by Meir and Schmidt
\cite{Meir2004}.

In the following, we will describe the mathematical basis of the 
integro-differential equation approach in the quasistatic 
approximation.
Then we will present the developed 
numerical model that utilizes the open source CFD library
OpenFOAM\TReg\ \cite{openfoam},  supplemented by an MPI-parallelized implementation of Biot-Savart's law. We will discuss, in particular, 
the convergence properties of this numerical scheme when
applied to a TI problem in cuboid geometry.
Based on this, we will study the effect of a 
varying geometric aspect ratio on the spatial structure of the TI 
and on the critical current. We will also discuss the 
effect of spontaneous 
chiral symmetry breaking as it was recently discussed by 
Gellert \etal \cite{Gellert2011} and 
Bonanno \etal \cite{Bonanno2012}. Then, we will apply 
our model to the cylindrical TI experiment \cite{Seilmayer2012}, for which we
will show a surprising agreement between measured and 
computed growth rates.
The paper closes with a discussion of the results, with an 
outlook towards more detailed simulations of liquid metal batteries, 
and with prospects for a larger TI-related experiment.

\section{Mathematical Model}
The initial point for describing fluid dynamics in a liquid metal 
is the Navier-Stokes equation (NSE)
\begin{eqnarray}\label{eqn:navierstokes}
\dot {\bi v} + \left({\bi v}\cdot\nabla\right){\bi v} = 
- \frac {\nabla p}{\rho} + \nu \Delta {\bi v} + \frac{\bi f }{\rho},
\end{eqnarray}
with ${\bi v}$ denoting the velocity, $p$ the pressure, 
$\rho$ the density, $\nu$ the kinematic viscosity, and ${\bi f}$ the 
body force density.  For incompressible fluids the continuity equation
$\nabla \cdot {\bi v} = 0$ has to be taken into account. 

The trigger of the TI is the Lorentz force,
\begin{eqnarray}
\bi f = \bi f_{\mathrm L} = \bi J \times \bi B,
\end{eqnarray}
with $\bi J$ meaning the current density. Note  that the magnetic 
field ${\bi B}$ consists of two parts: 
the static field $\bi B_{\mathrm 0}$, generated by the 
applied current $I$ (or the corresponding current density $\bi J_{\mathrm 0}$), 
and the induced magnetic field ${ \bi b}$, generated by the motion of the 
electrically conducing fluid.
Although in our problem ${ \bi b}$ is rather small, it must not 
be neglected in the expression for the Lorentz force 
because it is multiplied with the 
large current density ${\bi J_{\mathrm 0}}$.

The usual way to simulate the magnetic field evolution 
is by solving the induction equation
\begin{eqnarray}
\dot {\bi B} + ({\bi v}\cdot\nabla){\bi B} &= \left({\bi B}\cdot \nabla\right){\bi v}+
\frac{1}{\mu_0 \sigma}\Delta {\bi B}  \;.
\end{eqnarray}
The solution of this equation requires suitable 
boundary conditions for the magnetic field, 
which can be implemented either by solving a Laplace equation
in the exterior of the fluid 
\cite{Kluwer1999,Guermond2007}, or by equivalent boundary element 
methods \cite{Stefani2000,Iskakov2004,Giesecke2008}.

In the following, we choose a 
second option, i.e. we compute the magnetic field using  
Biot-Savart's law
\begin{eqnarray}\label{eqn:biotsavart}
{\bi B}({\bi r}) = \frac{\mu_0}{4\pi}\int dV' \, 
\frac{{\bi J}({\bi r}') \times ({\bi r}-{\bi r}')}{\left|{\bi r}-{\bi r}'\right|^3},
\end{eqnarray}
which is the inversion of Amp{\`e}re's law 
$\nabla \times \bi B = \mu_0 \bi J$. 
Here ${\bi r}$ means the point coordinate where $\bi B$ has to be computed, 
and ${\bi r}'$ is the integration coordinate, which 
runs through the whole domain where ${\bi J}$ exists.

This way, the problem is shifted from the determination of the 
magnetic field $\bi B$ to the determination of the current density
$\bi J$. 
The approach to use Biot-Savart's law was 
already proposed and utilized by 
Meir \etal \cite{Meir2004}, and will be adopted to our case.

\begin{figure}[hbt]
\centerline{
\includegraphics[width=8cm]{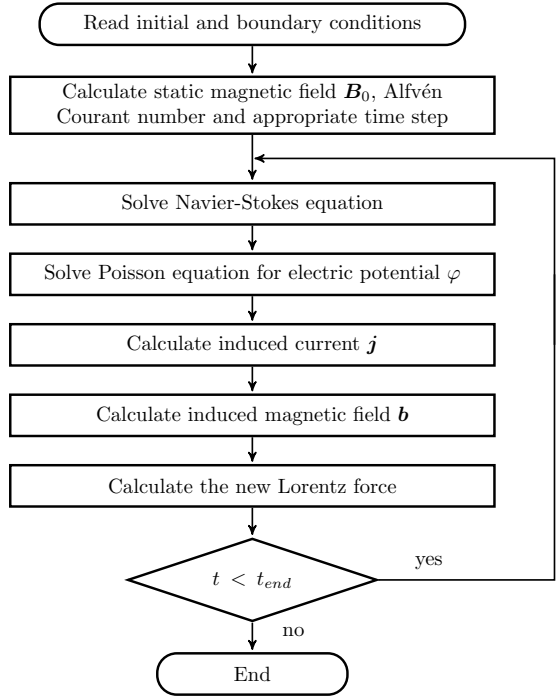}}
\caption{Flow chart of the simulation model.}
\label{fig:workflow}
\end{figure}

The flow chart of the numerical model is shown in figure
\ref{fig:workflow}. Assuming the current $I$ through our cylindrical
vessel as given (e.g. the battery charging current), we compute the 
corresponding current density $\bi J_{\mathrm 0}$, and the associated 
static magnetic 
field $\bi B_{\mathrm 0}$.  For an infinitely long 
vertical cylinder we would obtain
\begin{eqnarray}
{\bi B_{\mathrm 0}}(x,y) = \frac{\mu_0 J_0}{2}\left(y \bi {e_x}-x \bi{ e_y}\right),
\end{eqnarray}
in Cartesian coordinates $x,y,z$. 
For  real geometries with external leads to the system, $\bi B_{\mathrm 0}$ 
could still be be computed by Biot-Savart's law. 

In the main loop of our numerical scheme, 
the Navier-Stokes equation (\ref{eqn:navierstokes}) 
is solved firstly, followed by a velocity corrector step to ensure 
continuity ($\nabla \bi \cdot \, \bi v = 0$). 
Then we have to find the electric current density. Presupposing the
magnetic Reynolds number $Rm=\mu_0 \sigma R v$ on the basis of the
the TI-triggered  
velocity scale $v$ to be small, we can invoke the 
quasistatic approximation \cite{Davidson2001}. This means
that we express the electric field by the gradient of an 
electrostatic potential, ${\bi E}=- \nabla \Phi$.
Applying the divergence operator to Ohm's law in moving conductors,
${\bi J}=\sigma(-\nabla \Phi + {\bi v} \times {\bi B})$, 
and demanding charge conservation, $\nabla \cdot {\bi J}=0$, we 
arrive at the  Poisson equation 
\begin{eqnarray} \Delta\varphi = 
\nabla\cdot\left({\bi v} \times {\bi B}\right)
\end{eqnarray}
for the perturbed electric potential 
$\varphi = \Phi - J_0 z/\sigma$, where 
$z$ is the coordinate along the battery axis 
pointing into the direction of the applied 
potential difference.
By subtracting the electric potential caused by the battery charging 
current $I$ we can easily set the boundary conditions at the electrodes 
to
$\varphi = 0$. Since no current is allowed to flow through 
the insulating rim
of the cylindrical vessel, we assume here $\nabla \varphi=0$.

Having derived the potential in the fluid by solving the Poisson equation, 
we easily recover the current density induced by the fluid motion as
\begin{eqnarray}
{\bi j} = \sigma\left(-\nabla\varphi + {\bi v}\times {\bi B}\right)
\end{eqnarray}
and then the induced magnetic field by using Biot-Savart's law 
(\ref{eqn:biotsavart}). The total current density is
\begin{eqnarray}
{\bi J} = \bi j - J_0 \bi e_z.
\end{eqnarray}

In a last step, the Lorentz force, to be implemented into the 
Navier-Stokes equation, 
is calculated according to 
\begin{eqnarray}
\bi f_{\mathrm L} = \bi J\times {\bi B} = \left(\sigma\left(-\nabla\varphi + {\bi v}\times\left(\bi B_{\mathrm 0} + {\bi b}\right)\right) - J_0 \bi e_z\right)\times \left(\bi B_{\mathrm 0} + {\bi b}\right).
\end{eqnarray}
It should be noticed again that the (small) induced magnetic field 
$\bi b$ cannot be omitted here, since its product with the (large) 
impressed current $\bi J_0$ is of the same order
as the product of the (small) induced current $\bi j$ with the (large) 
magnetic field $\bi B_0$. 

Assuming an implementation of this scheme on a certain grid, one has to ask 
for the allowed time steps to keep the time evolution of the 
system stable. 
For this it is important to consider 
not only the Courant number based on the hydrodynamic velocity
$v$, but also the Alfv{\'e}n-Courant number based on the 
Alfv{\'e}n velocity $v_{A}=B/(\mu_0 \rho)^{1/2}$. 
 The influence of the 
Alfv{\'e}n-Courant number on the accuracy of the 
simulations is discussed  in the following section.

\section{Numerical scheme and convergence}
In this section we will examine the numerical scheme
with respect to its convergence and error characteristics.
We will both study variations of the grid spacing as well as
of the time steps. We will further consider the case that 
different grid spacings and time steps are used for the 
Navier-Stokes equation and for the Biot-Savart law. 

\begin{figure}[hbt]
\centerline{
\includegraphics[height=\hDia]{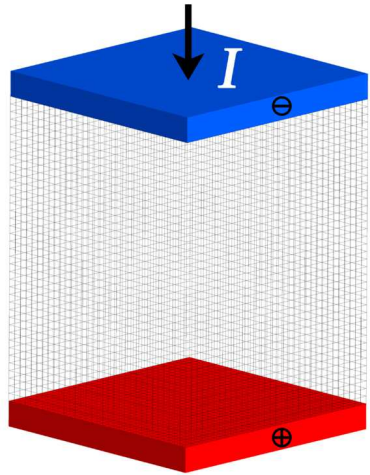}}
\caption{Geometric setting for the simulation 
of the TI in cuboid geometry. The electric current is 
assumed to be impressed by infinitely long electrodes.
The Navier-Stokes equation, enhanced by the Poisson equation 
and the Biot-Savart law, is solved in the gridded volume between
the electrodes.}
\label{fig:mesh3d}
\end{figure}

In order to utilize the convenience of orthogonal cells for grid 
refinement, a cuboid with 
the dimensions of $96\times96\times\SI{120}{mm^3}$ 
is used as a reference geometry, see figure \ref{fig:mesh3d}.

Anticipating the later simulation of  the recent 
TI experiment \cite{Seilmayer2012},
we will start to work here with the 
real material parameters
of \BPChem{GaInSn}: electric conductivity 
$\sigma=\SI{3.29e6}{\siemens\per\metre}$, 
density $\rho=\SI{6403}{\kilogram\per\metre^3}$, 
kinematic viscosity $\eta = \SI{3.4e-7}{\metre^2\per\second}$. 
The applied current is set to \SI{10}{\kilo\ampere}.

\begin{figure}[hbt]
\centerline{
\includegraphics[height=\hDia]{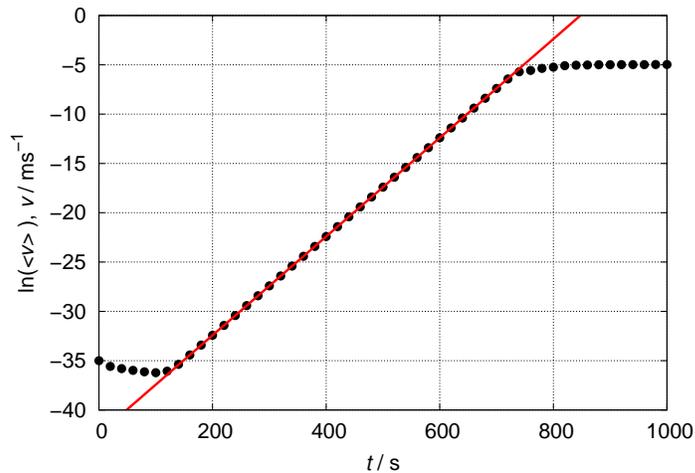}}
\caption{The typical flow development of the TI is characterized 
by an initial phase, an exponential 
growth phase, and a final saturation phase (black dotted line). 
The growth rate corresponds to the slope of the red line.}
\label{fig:gr_prinzip}
\end{figure}

For this case, figure \ref{fig:gr_prinzip} 
shows a typical evolution of the TI in time, 
comprising an initial
phase were the eigenfield of the TI develops, 
followed by a long period (over
 $\sim$ \SI{600}{\second}) in which this eigenfield 
 grows exponentially, and a
 final phase in which the TI 
 saturates at a certain velocity level.

In the following we will compare  the growth rates, 
as they can be read off
during the exponential growth phase 
for different grid spacings and time steps,
with respect to their relative errors
which are derived by dividing the absolute error by 
the difference of the TI 
growth rates at \SI{10}{\kilo\ampere} and \SI{0}{\kilo\ampere}.

\subsection{Grid refinement}
\begin{figure}[hbt]
\centerline{
\includegraphics[height=\hDia]{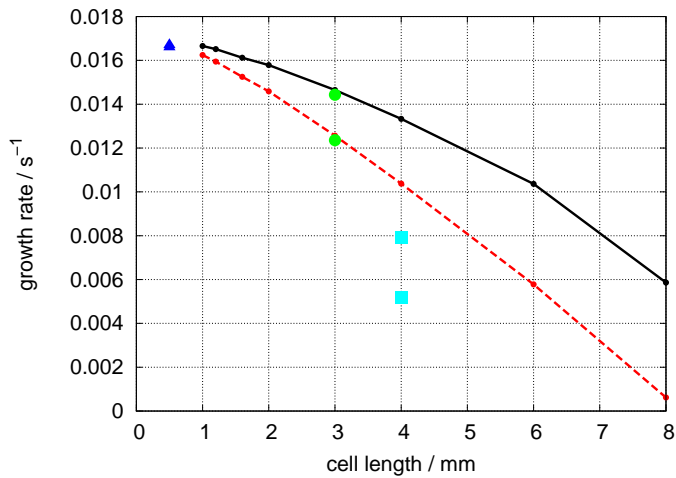}}
\caption{Influence of the grid cell size on the growth
  rate. Simulation is carried out for linear cell-face interpolation
  (red dashed line) as well as cubic one (black line). Computing
  Biot-Savart's law with $\SI{8}{\milli\metre}$ cells does not
  give good results ({\tiny $\textcolor{cyan1}{\blacksquare}$}), while a
  $\SI{6}{\milli\metre}$ grid is sufficient ({\large $\textcolor{green1}{\bullet}$}). This technique allows an additional refinement of the
  initial mesh, i.e. to solve the NSE on a
  $\SI{0.5}{\milli\metre}$ grid ($\textcolor{blue1}{\blacktriangle}$).}
\label{fig:grid}
\end{figure}

\begin{figure}[hbt]
\centerline{
\includegraphics[height=\hDia]{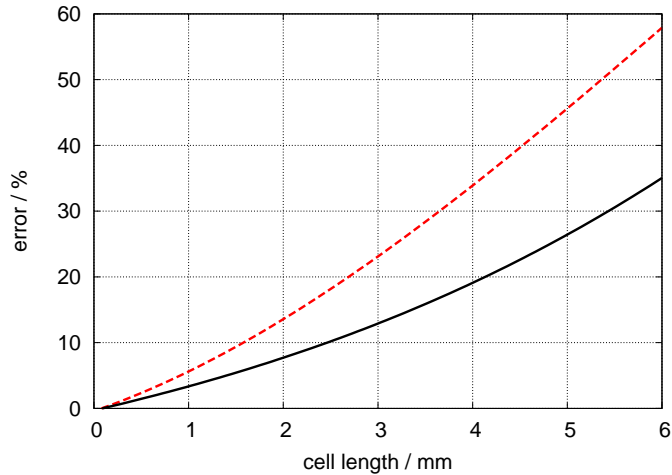}}
\caption{Influence of the grid cell size on the error of the growth rate. The maximal growth rate at \SI{0}{\milli\metre} was extrapolated from \SI{1}{\milli\metre} cells and used as reference. The black line stands for cubic cell-face interpolation, the red dotted one for linear interpolation.}
\label{fig:griderror}
\end{figure}

We start by exploring the effects of different grid sizes
of the cubic cells with 
an edge length between \SI{1}{\milli\metre} and \SI{8}{\milli\metre}. Using the
Finite-Volume-Method (FVM) of OpenFOAM\TReg\  
for the simulation, the values from the cell centres have to be
interpolated to the cell faces in order to ensure continuity. 
For this interpolation we employ two different methods, a linear one and
a cubic one, and compare the results in figure \ref{fig:grid}. As 
expected, the cubic interpolation provides better results, i.e. the 
growth rates are typically higher. While it would be numerically very 
demanding to go 
below a cell size of \SI{1}{mm}, we expect the difference between
cubic and linear interpolation to be zero in the limit of 
vanishing cell size. 
With that assumption we do a Richardson extrapolation to 
obtain the ''true'' growth rate of the TI. Taking this value  
as reference, 
the relative error of the 
growth rates is then calculated in dependence on 
the grid cell size (figure \ref{fig:griderror}).

Since the solution of the Biot-Savart law is numerically 
very costly, we have
additionally checked the possibility 
to speed up this part of the simulation. 
While the two lines in figure \ref{fig:grid} result from
computations in which the Navier-Stokes equation and the 
Biot-Savart law are implemented on the same grid,
the green circles and the liqht blue squares 
correspond to cases in which the Biot-Savart law is 
realized on coarser grids. 
In contrast to the ${ \bi b}$-calculation with cells of 
$\SI{8}{\milli\metre}$ edge length, which
does not give good results (see figure \ref{fig:grid}), 
a grid cell size of $\SI{6}{\milli\metre}$
already seems to be sufficient. Solving the 
Navier-Stokes equation on a
$\SI{3}{\milli\metre}$ grid, and 
computing Biot-Savart's law on a
$\SI{6}{\milli\metre}$ grid, provides almost the 
same results as doing both on a
$\SI{3}{\milli\metre}$ grid. 
This is numerically very convenient because Biot-Savart's law
needs the square of cell numbers as operations whenever 
carried out, i.e. a
coarse grid makes simulation much faster. The resulting slightly lower 
growth rates when working with two meshes appear due to the inter-grid mapping. 
The interpolation flattens the ${ \bi b}$-field a little bit 
which reduces the growth rate -- but the difference is very small. 

\subsection{Alfv{\'e}n-Courant number and time steps}
\begin{figure}[hbt]
\centerline{
\includegraphics[height=\hDia]{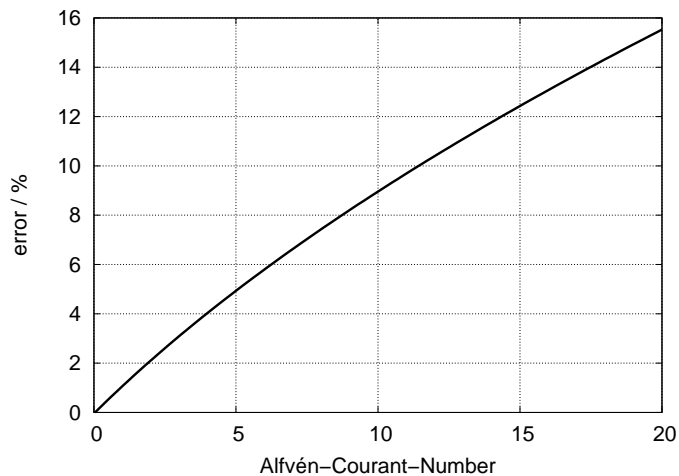}}
\caption{Influence of the Alfv{\'e}n-Courant number on the error of the growth rate. The Alfv{\'e}n-Courant number is one, if a particle, moving with Alfv{\'e}n velocity, passes exactly one grid cell in one time step.}
\label{fig:AlfvenCoNr}
\end{figure}

As usually, during the time integration of the NSE we have to 
fulfill the Courant-Friedrichs-Lewy (CFL) condition for the 
time step. 
In our special problem, however, information is not only
transferred by the flow velocity, but also electromagnetically with
the Alfv{\'e}n velocity $v_A = B/(\mu_0 \rho)^{1/2}$.
We consider this by introducing an ``Alfv{\'e}n-Courant number'' and 
compute the maximum time step on that basis. Due to the high currents at 
which TI is expected to set in, and the
consequently high magnetic field $B_0$, this time step 
has to be very
short in spite of the low fluid velocities. 
Fortunately, it turns out that choosing the
Alfv{\'e}n-Courant number larger than one does not lead 
to a completely unstable system, as it would be expected for a purely
differential equation system. Evidently, the mixed integro-differential
equation approach makes the system somewhat more ``benign'' and
less vulnerable to violations of the CFL condition.
Of course, the error increases
significantly with increasing Alfv{\'e}n-Courant number as shown in figure 
\ref{fig:AlfvenCoNr}.

\begin{figure}[hbt]
\centerline{
\includegraphics[height=\hDia]{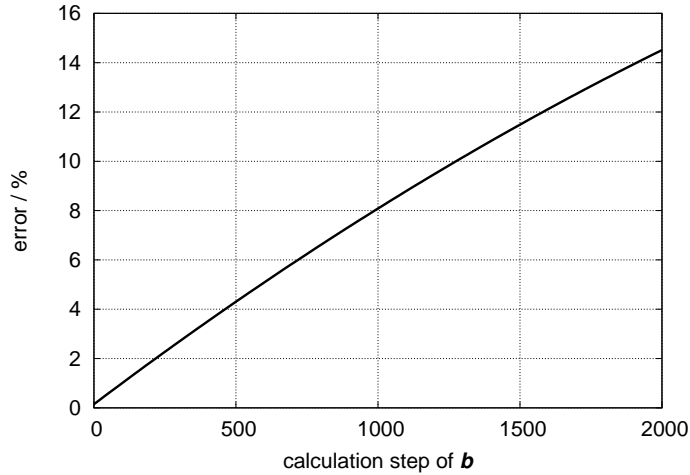}}
\caption{Error of the growth rate of the TI depending on how often the induced magnetic field $\bi b$ is calculated. As the TI is growing slowly and therefore $\bi b$ changes slowly, the simulation can be accelerated by doing Biot-Savart not every time step.}
\label{fig:bStep}
\end{figure}

Besides calculating the magnetic field ${ \bi b}$ on a coarser grid than
the velocity field, we have also tested
to compute it only at every n\textsuperscript{th}  timestep. Figure \ref{fig:bStep} 
shows that it is well possible to compute the
Biot-Savart integrals only every 100\textsuperscript{th} 
time step.  The benefit in 
computation time is especially high when solving 
the NSE and Biot-Savart's law on the same grid, as 
the latter one takes usually more than 100 times 
longer than solving all other equations.

\section{Results for cuboid geometry}
In this section we will focus on the characteristic 
eigenmode structure of the TI, on the typical 
velocities in the saturated regime, 
and on the critical currents in dependence 
on various geometry parameters. 
The simulation are done for a cuboid volume with a 
base area of $96\times 96$ mm$^2$, but with varying heights.

\subsection{Shape, alignment, and chiral symmetry breaking}
\label{ch:helix}
\begin{figure}[hbt]
\centerline{
\includegraphics[height=7cm]{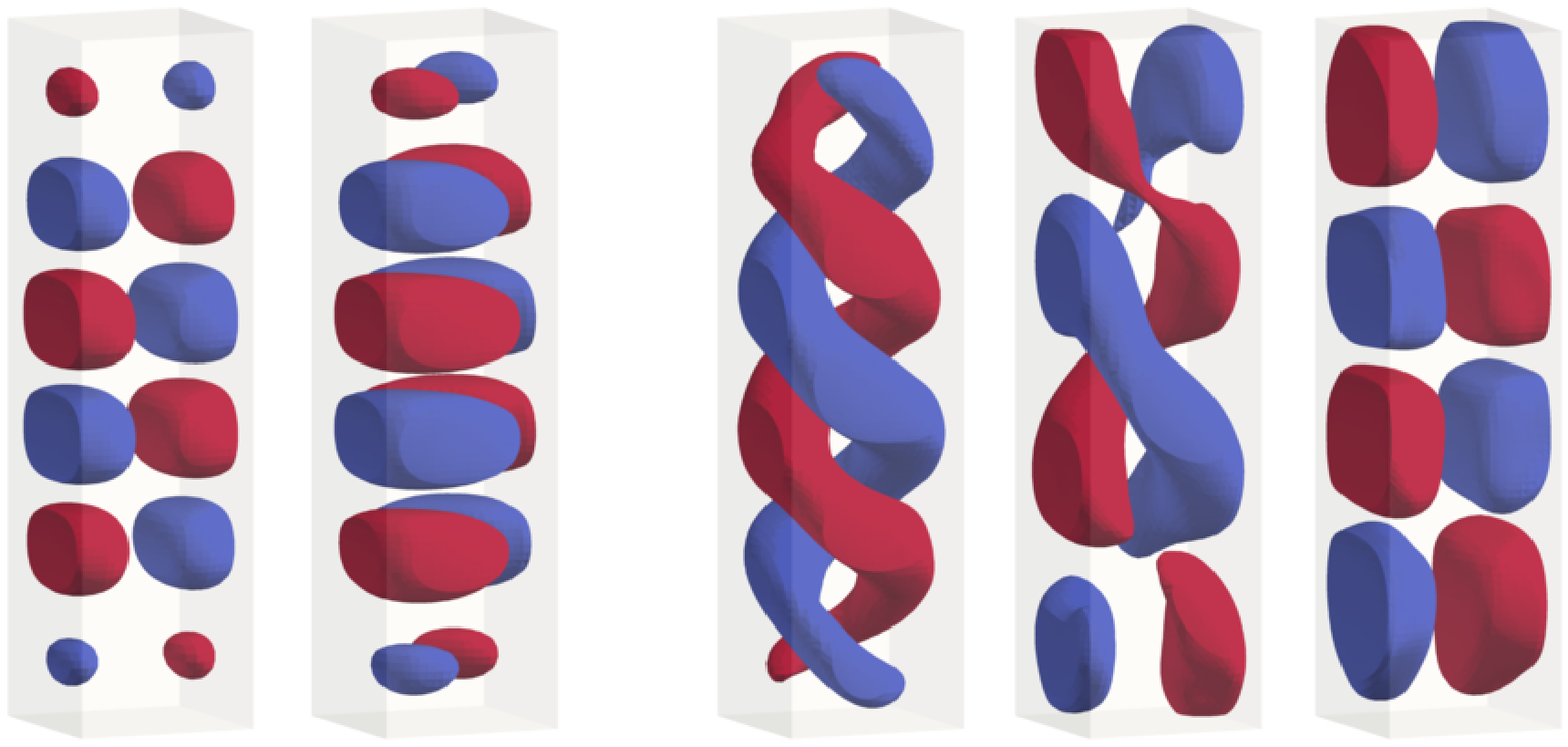}}
\caption{Iso-surface plots of the magnetic field 
$b_z$ for a cuboid of $h=\SI{360}{\milli\metre}$. 
Depending on the initial velocity conditions, the 
convection cells of the TI align to an arbitrary 
direction (left: orthogonal, second: diagonal to base cube).

Due to a helical perturbation, a helical TI may 
grow (3. panel, $t=\SI{2300}{s}$), but almost 
disappears later (4. panel 
 $t=\SI{2400}{s}$, right $t=\SI{2800}{s}$). The 
 corresponding time graph is shown in figure \ref{fig:helix}.
}
\label{fig:shape}
\end{figure}

Basically, the TI appears in form of convection cells 
that are
vertically stacked one over another (see figure 
\ref{fig:shape}). The fluid in one 
cell rotates around some horizontal axis, and  
the sense of rotation is changing from one cell to 
the neighboring one. Referring, for the moment, to cylindrical geometry,
this cell structure represents a non-axisymmetric flow
structure, with an azimuthal wavenumber
$m = \pm 1$. Since the present simulations are carried out 
for cuboid geometry, one could imagine, e.g., 
a preferred alignment of the convection cells 
with the diagonal of the base area. However, this is generally 
not 
the case. In fact, the alignment of the cells depends 
strongly on the chosen initial conditions. Assuming a small 
perturbation in the $x$-direction, the TI will arise 
around the $y$-axis (figure \ref{fig:shape}). 
For arbitrary initial velocity, the TI will develop 
around a random axis.

\begin{figure}[hbt]
\centerline{
\includegraphics[height=\hDia]{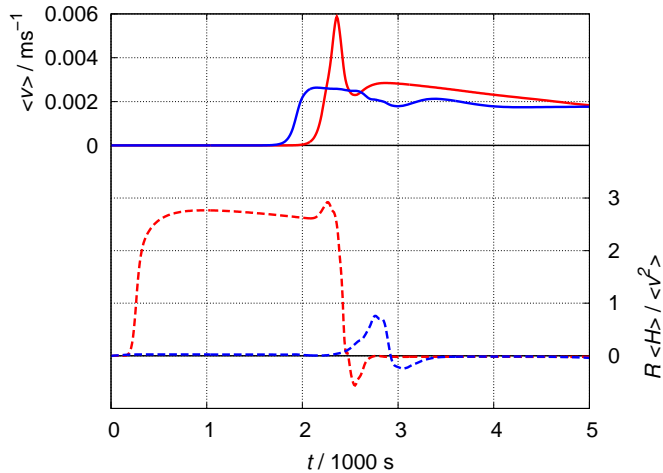}}
\caption{Temporal evolution of the mean velocity 
(continuous lines in the upper part) and of the 
corresponding normalized helicity 
(dotted lines in the lower part), 
with $H=\bi v \cdot (\nabla \times \bi v)$ and 
$R$ as the half side length 
of the base square. The height of the cuboid 
is \SI{360}{\milli\metre}, the applied current 
\SI{10}{\kilo\ampere}. The blue lines correspond 
to an initial velocity along the diagonal of the 
base cube, while the red plots result from an 
initial velocity in $z$ direction.}
\label{fig:helix}
\end{figure}

Most interestingly, a transient helical shape of the TI may result
under certain  initial conditions
(see figure \ref{fig:shape}, right). This was 
observed, e.g., with an initial velocity in $z$-direction 
and in particular when the height of the simulated geometry did 
not match an integer multiple of the wave length of 
the TI. The helical perturbation may develop if, e.g., 
two convection cells arise simultaneously, but with 
azimuthal twisted rotation axes. The growth rates of 
the non-helical and the helical TI state were 
found to be very 
similar. Figure \ref{fig:helix} shows the velocity 
and the normalized mean value  
of the helicity $H={\bi v} \cdot (\nabla \times {\bi v})$
of a non-helical (blue) and of a helical (red) state 
for a $96\times 96 \times 360$ mm$^3$ geometry 
($h=3.1\lambda$). Reaching a certain velocity level
in the saturated regime, 
the helicity of the helical state 
reduces significantly and goes finally to zero.
Approximetaly at the same time (at $\sim$2500 s) 
the saturated state of the formerly non-helical 
state acquires also some helicity, which goes 
finally to zero, too.
The ultimate convection cells of the two states are not 
equal (compare figure \ref{fig:shape} left 
and right), and seem to depend on the different 
initial conditions of the velocity.

This transient appearance of helical TI structures is 
quite remarkable, 
since it corresponds to the spontaneous chiral symmetry 
breaking which was recently observed by 
Gellert \etal \cite{Gellert2011}
and discussed in detail by Bonanno \etal 
\cite{Bonanno2012}. 
Here we can only evidence its occurrence, while
a more comprehensive study of this effect 
has to be left for future studies.

\subsection{Saturation levels of the TI}
\begin{figure}[hbt]
\centerline{
\includegraphics[height=\hDia]{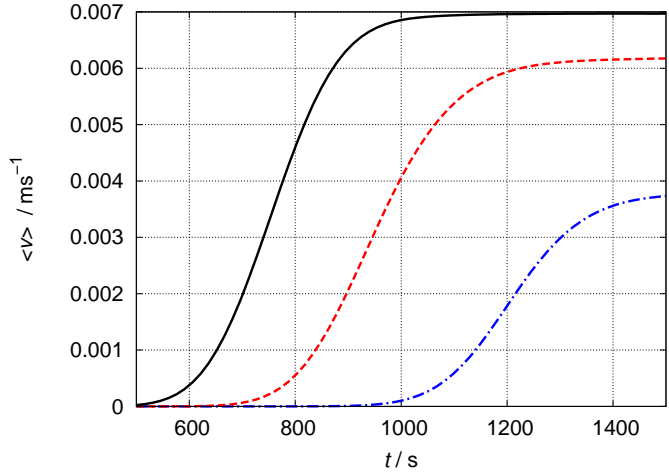}}
\caption{Development of the mean velocities 
of the TI in function of time for different 
applied currents: $I=\SI{14}{\kilo\ampere}$ 
(blue line), $I=\SI{16}{\kilo\ampere}$ 
(red line) and $I=\SI{18}{\kilo\ampere}$
(black line).}
\label{fig:u}
\end{figure}

\begin{figure}[hbt]
\centerline{
\includegraphics[height=\hDia]{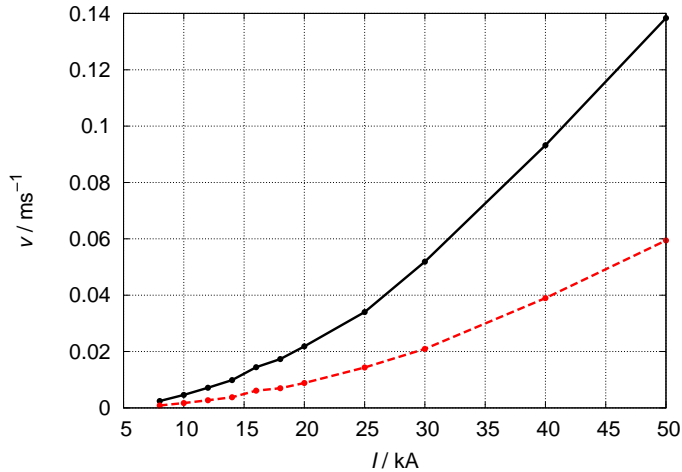}}
\caption{Dependence of velocity of the TI on 
the applied current $I$. The black line stands 
for maximum velocities, the red dashed line 
for averaged velocities.}
\label{fig:vMax}
\end{figure}

To investigate the saturation level 
of the TI we have carried out simulations 
using different currents 
for a cuboid geometry 
of 96$\times$96$\times$120\,mm$^3$  
and a grid cell length of \SI{3}{\milli\metre}. When 
the critical 
current for the onset of the TI is exceeded, any small 
disturbance 
can initiate the instability which then 
needs some time to establish its eigenmode. 
Thereafter, this eigenmode grows exponentially, 
see figure \ref{fig:u}. For currents only slightly 
above the critical one, the instability grows very slowly. 
In such a case, it may take tens of minutes until the TI 
reaches its saturated final state. Figure \ref{fig:vMax} 
shows the dependence of the final velocities on the 
applied current.  We see that the maximum velocities 
grow from a few mm/s close to the critical current, to nearly 
14\,cm/s at \SI{50}{\kilo\ampere}. 
Velocities on this scale would clearly be
problematic for the stability of the layer
stratification in liquid metal batteries. 

\subsection{Critical current in dependence on the aspect ratio}

To investigate  the dependence of 
the critical current on the aspect ratio of the
fluid volume we consider  a 
cuboid with a base area of $96\times\SI{96}{\milli\metre^2}$ 
which is changed in height from 24 to \SI{480}{\milli\metre}. 

As a first estimate for the critical wavelength of the TI
we can take the value for an infinitely long cylinder 
of radius $R$,
\begin{eqnarray}
\lambda = 2\pi R / k,
\end{eqnarray}
for which the critical wavenumber $k$ is known to 
be 2.47 \cite{Ruediger2012b}. Embedding 
this cylinder in the cuboid to be considered
here, the resulting  wave length would be 
$\lambda = \SI{122}{\milli\metre}$. 
To obtain the corresponding true spatial wave length for 
the cuboid geometry, the TI is simulated for a very long 
cuboid with a 
height of \SI{1040}{\milli\metre}. For this case we obtain 
$\lambda=\SI{115}{\milli\metre}$, a value quite close to 
the corresponding value of the infinitely long cylinder.

\begin{figure}[hbt]
\centerline{
\includegraphics[height=\hDia]{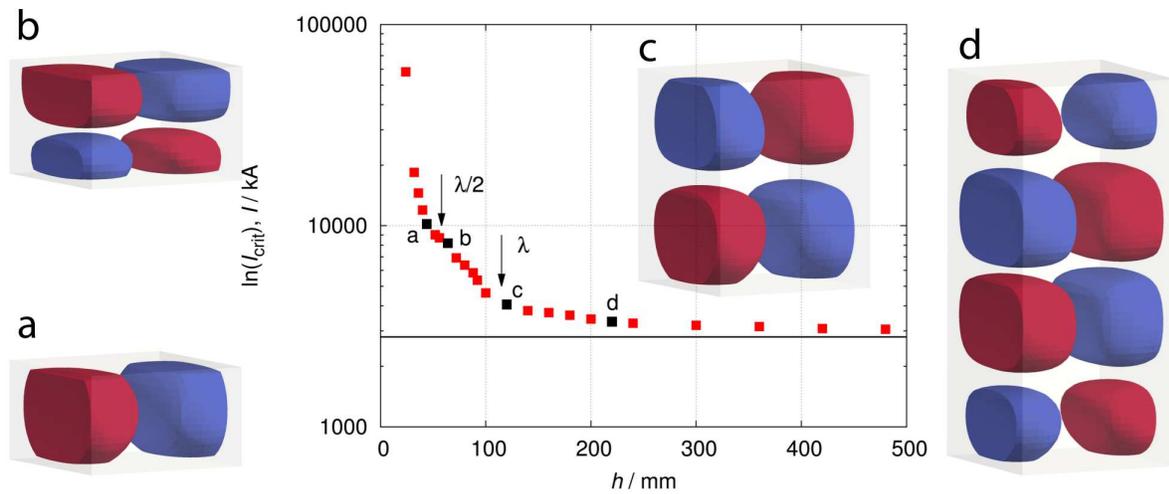}}
\caption{Dependence of the critical current for 
the onset of the TI on the height 
of a cuboid of 
$96 \times 96 \times h$ mm$^3$. 
If the height of the vessel drops below the wave 
length of the TI, the critical current increases 
steeply. 
The black line marks the critical current for the 
infinite cylinder.

The isosurface plots show the induced magnetic 
field ${b_z}$ for four selected heights $h$, 
indicated by the  
black cubes in the graph. Below one half wave 
length ($h<\lambda / 2$), 
only one convection cell is present (a). 
Slightly above $h=\lambda / 2$, there 
are two cells (b). Panel c shows the case when $h = \lambda$. 
For $h=1.9\lambda$ (d), we observe four cells,
the top and bottom ones appearing 
somewhat smaller.}
\label{fig:h}
\end{figure}

Figure \ref{fig:h} shows now the critical currents 
as a function of height. In the limit of very tall 
cuboids, the critical current converges to a
value of approximately \SI{3}{\kilo\ampere} 
which is also quite close to the
value \SI{2.7}{\kilo\ampere} for  the infinite long 
cylinder \cite{Ruediger2012b}. Shortening the cuboid, 
a first plateau of the critical current is observed 
at a height corresponding to the 
TI wave length of $\SI{115}{\milli\metre}$. 
Decreasing the height further,
a sharp increase of the critical current is observed, 
which flattens again to a second plateau
when the height comes close to the half wave length
($\SI{57.5}{\milli\metre}$). Shortening the cuboid 
still further, the critical currents rise extremely steep.
With regard to liquid metal batteries this would mean
that very flat cells will not be affected by the
Tayler instability.
 
For those heights which are below the half wavelength of 
the TI ($h<\lambda / 2$), there is only one 
convection cell, for slightly taller vessels 
a second one appears (see figure \ref{fig:h}, left). 
In general, the number of convection cells adapts to the
height, and the cell size acquires a maximum if the height is a
multiple of the wavelength. 
If $h$ exceeds $1.5\lambda$, a third 
cell may appear -- but only in case of a helical TI. In 
case of non-helical TI an even number of convection 
cells seems to be more stable. Usually, the new convection 
cells appear at the top and/or bottom. If the height is 
slightly below an integer multiple of the characteristic 
wave length, the upper and lower convection cell will appear 
somehow ''compressed'' in comparison to 
the more central ones (figure \ref{fig:h}, right).

\section{Validation of experimental results}\label{ch:experiment}

In a recent paper \cite{Seilmayer2012}, we had evidenced the
occurrence of the TI in a liquid metal experiment. 
A cylinder, made of 
polyoxylmethylene ($d=\SI{100}{\milli\metre}$, $h=\SI{750}{\milli\metre}$) 
was filled with the eutectic \BPChem{GaInSn}, being 
liquid at room temperature. A current of up to 8 kA 
was injected into the liquid metal 
through two massive copper electrodes 
at the top and bottom of the cylinder. 
In order to avoid any inhomogeneities of the
electric current at the interface 
between Cu and GaInSn, we restrained from inserting 
ultrasonic transducers for the direct measurement of 
the velocity perturbations, and relied
exclusively on external measurements of the induced magnetic field. 
The vertical component $b_z$  was measured at 
11 positions along the vertical axis of the device.
With this set-up we were able to identify the growth rates
of the short-wavelength perturbations due to the TI, before 
those gave way to longer wavelength structure that we
attributed to the thermal wind resulting from Joule 
heating in the GaInSn.  

For this cylindrical geometry, we simulate the TI and compare
the results with the available experimental data.
The cylindrical fluid volume
is meshed with a grid cell size of 
about $3\times 3\times 3$ mm$^3$. 
As the electrodes are relatively 
thick and are elongated by long power supply leads, we 
assume 
$\bi B_{\mathrm 0}$ to be constant over the height of the cylinder.

\begin{figure}[hbt]
\centerline{
\includegraphics[width=14cm]{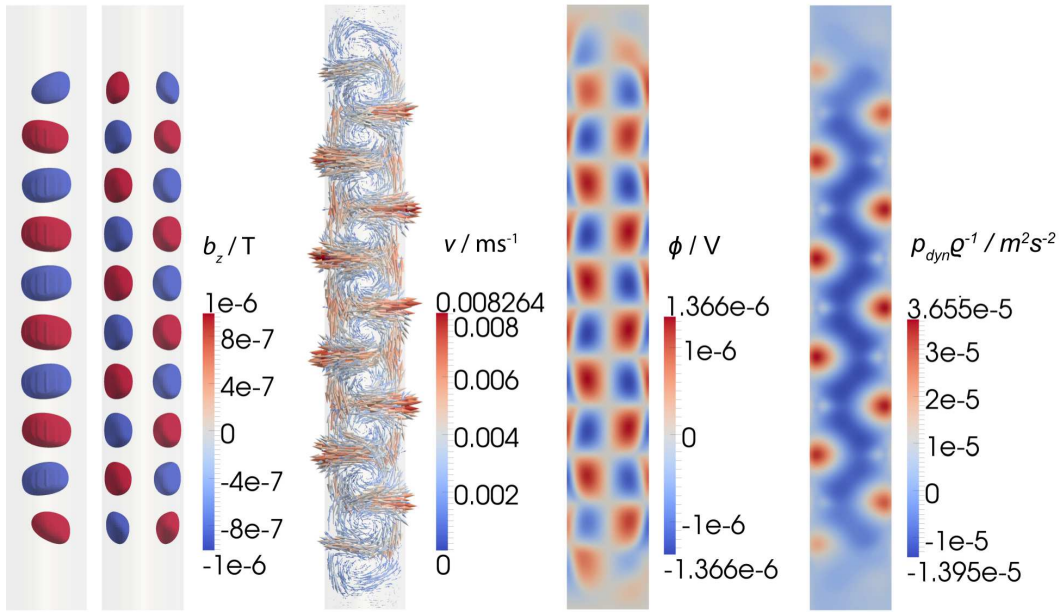}}
\caption{Iso-shapes for $b_z$ (left), velocity 
field (second panel), electric potential (third panel) and dynamic 
pressure (right) of the saturated Tayler instability in a cylinder filled with \BPChem{GaInSn}. The applied current is \SI{10}{\kilo\ampere}. The magnetic field is shown in $x$ and $y$ direction, all other figures only in $x$.}
\label{fig:walzen}
\end{figure}

For an assumed current of \SI{10}{\kilo\ampere}, the resulting 
convection cells 
of the TI are illustrated in figure \ref{fig:walzen}.
It should be noticed that the number and shape of these 
convection cells is quite independent on the initial 
conditions.

\begin{figure}[hbt]
\centerline{
\includegraphics[height=\hDia]{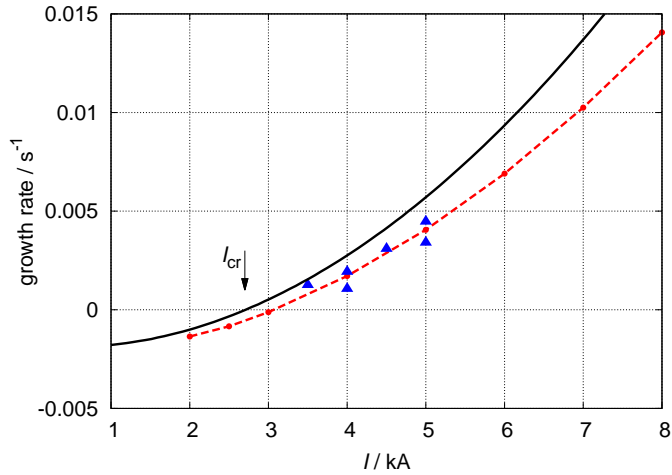}}
\caption{Growth rates of the Tayler instability 
in a cylinder filled with \BPChem{GaInSn}. The 
black line marks the maximal possible growth 
rates for the infinite cylinder 
\cite{Ruediger2012b}. Our simulation (red dashed line) fits well the experimental data (blue triangles) \cite{Seilmayer2012}.}
\label{fig:gr}
\end{figure}

The experimental growth rates are compared with the 
numerical ones in  figure \ref{fig:gr}. 
Simulations for the infinitely long
cylinder 
\cite{Ruediger2012b} 
provide the maximum possible growth rates. As we do simulation for the real, i.e. 
finite cylinder, the resulting growth rates lie slightly below 
these values, but correspond nicely with the experimental data.

\begin{figure}[hbt]
\centerline{
\includegraphics[height=\hDia]{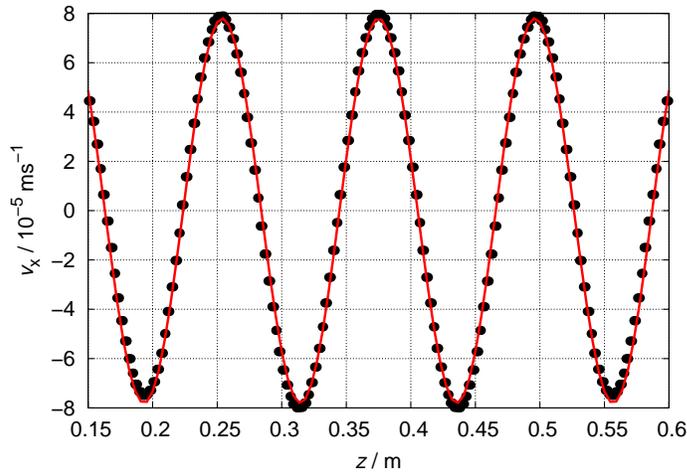}}
\caption{Velocities of the TI in $x$-direction 
along the $z$-axis in the central part. Values are taken at 
maximal growth rate, i.e. shortly before 
reaching saturation (black dotted line). 
The red fit allows to determine the spatial 
wavelength of $\lambda = \SI{121}{\milli\metre}$.}
\label{fig:lambdaExp}
\end{figure}

For our geometry 
the theoretical wave length is
$\lambda = \SI{127}{\milli\metre}$ \cite{Ruediger2012b}. This 
value matches fairly well the wave length 
obtained by our finite length 
simulation (\SI{121}{\milli\metre}), 
as inferred from figure \ref{fig:lambdaExp}.

\section{Conclusion and outlook}
On the basis of an integro-differential equation approach we 
have developed a numerical tool that is particularly
suited for investigations of current driven magnetohydrodynamic
instabilities in liquid metals. 
Employing the quasistatic approximation,
the solution of the 
induction equation for the magnetic field was substituted by 
solving a Poisson equation for the induced current density, 
from which
the magnetic field is computed via Biot-Savart's law. 
This procedure allows to deal with instabilities in 
conducting fluids characterized by
very small magnetic Prandtl numbers.

After examining the convergence properties of this numerical
scheme, 
we have simulated the TI in different 
geometries, including cuboids and cylinders. For the cuboid
it was shown in detail how the critical current 
increases with decreasing aspect ratio, exhibiting two 
remarkable plateaus 
at those heights that correspond to one 
or a half wavelength of the TI mode, respectively. 
This curve may be directly used for defining a 
maximum aspect ratio of liquid metal batteries at a 
given charging current. Depending on the initial 
conditions our simulations have also
shown the transient appearance of helical structures, 
i.e. the occurrence of chiral symmetry breaking.

For cylindrical geometry, we have numerically 
reproduced the growth rates 
of the TI that had been measured in a recent liquid metal 
experiment.

A main result of our simulations is 
that the saturation level
of the TI induced velocity for typical electrical 
currents are in the order of mm/s until cm/s. This is quite 
consistent with a formal scaling, by $\nu/R$, of the 
velocity levels as obtained at much larger magnetic Prandtl 
numbers \cite{Gellert2009}. While this good correspondence 
confirms the 
interaction parameter $\mathit{Ha}^2/Re$ to be the
governing parameter for the saturation over a wide 
range of magnetic Prandtl numbers,
it makes the physical interpretation of 
the saturation mechanism a bit more 
difficult. Actually, for $Pm$ close to one, Gellert \etal
\cite{Gellert2009} had identified a very plausible saturation
mechanism in terms of a radially dependent
increase of the TI triggered turbulent 
resistivity ($\beta$ effect)
that modifies the radial current distribution 
in such a way as to make the magnetic field distribution
just marginally stable. Evidently, this nice picture
does not apply any more for small $Pm$, as considered here,
since the arising velocity perturbation are much too weak
for producing any noticeable turbulent resistivity. 
In this respect, a simple picture for understanding
the saturation of TI at low $Pm$ remains elusive.

In a next step we plan to include 
into the numerical model the effects of 
thermal convection resulting from the Joule heating
due to the strong current, 
which has turned out important in the experiment. We also  
develop a multiphase solver which will allow  us 
to simulate three-layer liquid metal batteries in 
detail.

On the experimental side, in the framework of the DRESDYN project 
\cite{Dresdyn2012} we plan to set-up a large scale 
liquid sodium experiment for the combined investigation of the
magneto-rotational instability,
as recently investigated \cite{Stefani2009}, and the TI.

\section*{Acknowledgment}

This work was supported by Helmholtz-Gemeinschaft Deutscher
Forschungszentren (HGF) in frame of the ''Initiative f\"ur mobile und
station\"are Energiespeichersysteme'', and in frame of the 
Helmholtz Alliance LIMTECH, as well as
by Deutsche Forschungsgemeinschaft
in frame of the SPP 1488 (PlanetMag). We gratefully 
acknowledge fruitful discussions with
Rainer Arlt, Alfio Bonanno,
Marcus Gellert, Rainer Hollerbach,
G\"unther R\"udiger and Martin Seilmayer
on several aspects of the Tayler instability.

\section*{References}
\providecommand{\newblock}{}


\begin{thebibliography}{10}
\expandafter\ifx\csname url\endcsname\relax
  \def\url#1{{\tt #1}}\fi
\expandafter\ifx\csname urlprefix\endcsname\relax\def\urlprefix{URL }\fi
\providecommand{\eprint}[2][]{\url{#2}}

\bibitem{Seilmayer2012}
Seilmayer M, Stefani F, Gundrum T, Weier T, Gerbeth G, Gellert M and
  R{\"u}diger G 2012 {\em Phys. Rev. Letters\/} {\bf 108} 244501

\bibitem{Tayler1960}
Tayler R~J 1960 {\em Rev. Mod. Phys.\/} {\bf 32} 907 -- 913

\bibitem{Bergerson2006}
Bergerson W~F, Forest C~B, Fiksel G, Hannum D~A, Kendrick R, Sarff J~S and
  Stambler S 2006 {\em Phys. Rev. Lett.\/} {\bf 96} 015004

\bibitem{Vandakurov1972}
Vandakurov Y~V 1972 {\em Astron. Zh+.\/} {\bf 49} 324 -- 333

\bibitem{Tayler1973}
Tayler R~J 1973 {\em Mon. Not. R. astr. Soc.\/} {\bf 161} 365 -- 380

\bibitem{Spruit2002}
Spruit H~C 2002 {\em Astron. Astrophys.\/} {\bf 381} 923 -- 932

\bibitem{Zahn2007}
Zahn J~P, Brun A~S and Mathis S 2007 {\em Astron. Astrophys.\/} {\bf 474} 145
  -- 154

\bibitem{Suijs2008}
Suijs M~P~L, Langer N, Poelarends A~J, Yoon S~C, Heger A and Herwig F 2008 {\em
  Astron. Astrophys.\/} {\bf 481} L87 -- L90

\bibitem{Yoon2012}
Yoon S~C, Dierks A and Langer N 2012 {\em Astron. Astrophys.\/} {\bf 542} A113

\bibitem{Moll2008}
Moll R, Spruit H~C and Obergaulinger M 2008 {\em Astron. Astrophys.\/} {\bf
  492} 621 -- 630

\bibitem{Spies1988}
Spies G~O 1988 {\em Plasma Phys. Control. Fusion\/} {\bf 30} 1025 -- 1037

\bibitem{Cochran1993}
Cochran F~L and Robson A~E 1993 {\em Phys. Fluids B\/} {\bf 5} 2905 -- 2908

\bibitem{RUEDIGER2011}
R{\"u}diger G, Schultz M and Gellert M 2011 {\em Astron. Nachr.\/} {\bf 10} 1
  -- 7

\bibitem{Huggins2010}
Huggins R~E 2010 {\em Energy Storage\/} (Springer Sciene+Business Media)

\bibitem{Bradwell2012}
Bradwell D~J, Kim H, Sirk A~H~C and Sadoway D~R 2012 {\em J. Am. Chem. Soc.\/}
  {\bf 134} 1895 -- 1897

\bibitem{Weaver1962}
Weaver R~D, Smith S~W and Willmann N~L 1962 {\em J. Electrochem. Soc.\/} {\bf
  109} 653 -- 657

\bibitem{Cairns1969}
Cairns E~J and Shimotake H 1969 {\em Science\/} {\bf 164} 1347 --1355

\bibitem{Stefani2011}
Stefani F, Weier T, Gundrum T and Gerbeth G 2011 {\em Energ. Convers.
  Manage.\/} {\bf 52} 2982 -- 2986

\bibitem{Ruediger2012b}
R{\"u}diger G, Gellert M, Schultz M, Strassmeier K~G, Stefani F, Gundrum T,
  Seilmayer M and Gerbeth G 2012 {\em Astrophys. J.\/} {\bf 755} 181

\bibitem{Davidson2001}
Davidson P~A 2001 {\em An Introduction to Magnetohydrodynamics\/} (Cambridge
  University Press)

\bibitem{Meir2004}
Meir A~J, Schmidt P~G, Bakhtiyarov S~I and Overfelt R~A 2004 {\em J. Appl.
  Mech.\/} {\bf 71} 786 -- 795

\bibitem{openfoam}
{OpenFOAM Foundation} 2012 http://www.openfoam.org/

\bibitem{Gellert2011}
Gellert M, R\"udiger G and Hollerbach R 2011 {\em Mon. Not. R. Astron. Soc.\/}
  {\bf 414} 2696 -- 2701

\bibitem{Bonanno2012}
Bonanno A, Brandenburg A, {Del Sordo} F and Mitra D 2012 {\em Phys. Rev. E\/}
  {\bf 86} 016313

\bibitem{Kluwer1999}
Stefani F, Gerbeth G and Gailitis A 1999 Velocity profile optimization for the
  {R}iga dynamo experiment {\em Transfer Phenomena in Magnetohydrodynamics and
  Electrocinducting flows\/} ed Alemany A, Marty P and Thibault J~P (Kluwer
  Dordrecht) pp 31 -- 44

\bibitem{Guermond2007}
Guermond J~L, Laguerre R, L{\'e}orat J and Nore C 2007 {\em J. Comput. Phys.\/}
  {\bf 221} 349 -- 369

\bibitem{Stefani2000}
Stefani F, Gerbeth G and R{\"a}dler K~H 2000 {\em Astron. Nachr.\/} {\bf 321}
  65 -- 73

\bibitem{Iskakov2004}
Iskakov A, Descombes S and Dormy E 2004 {\em J. Comput. Phys.\/} {\bf 197} 540
  -- 554

\bibitem{Giesecke2008}
Giesecke A, Stefani F and Gerbeth G 2008 {\em Magnetohydrodynamics\/} {\bf 44}
  237 -- 252

\bibitem{Gellert2009}
Gellert M and R{\"u}diger G 2009 {\em Phys. Rev. E\/} {\bf 80} 046314

\bibitem{Dresdyn2012}
Stefani F, Eckert S, Gerbeth G, Giesecke A, Gundrum T, Steglich C, Weier T and
  Wustmann B 2012 {\em Magnetohydrodynamics\/} {\bf 48} 103 -- 113

\bibitem{Stefani2009}
Stefani F, Gerbeth G, Gundrum T, Hollerbach R, Priede J, R{\"u}diger G and
  Szklarski J 2009 {\em Phys. Rev. E\/} {\bf 80} 066303

\end{thebibliography}
\end{document}